\documentclass{elsart41}
\usepackage{graphics}
\usepackage{graphicx}
\usepackage{amssymb}
\begin{document}

\begin{frontmatter}



\title{Dynamics of quantum spin chains and multi-fermion excitation continua}
%

\author[ua]{O. Derzhko\corauthref{Derzhko}},
\ead{derzhko@icmp.lviv.ua}
\author[ua]{T. Krokhmalskii},
\author[de]{J. Stolze},
\author[us]{G. M\"{u}ller}

\address[ua]{Institute for Condensed Matter Physics NASU, 
1 Svientsitskii Str., L'viv-11, 79011, Ukraine}  
\address[de]{Institut f\"{u}r Physik,
Universit\"{a}t Dortmund, 44221 Dortmund, Germany}
\address[us]{Department of Physics, University of Rhode Island, 
Kingston, Rhode Island 02881-0817, USA}

\corauth[Derzhko]{Corresponding author. 
Tel: +38 0322 761978,
fax: +38 0322 761158}

\begin{abstract}

We use the Jordan-Wigner representation 
to study dynamic quantities 
for the spin-$\frac{1}{2}$ $XX$ chain in a transverse magnetic field.
We discuss in some detail 
the properties of the four-fermion excitation continuum 
which is probed by the dynamic trimer structure factor.

\end{abstract}

\end{frontmatter}


Recently
the subject of multi-magnon excitations of quasi-one-dimensional quantum spin systems 
has attracted considerable interest.
With high-resolution inelastic neutron scattering experiments one may expect 
to examine not only the bound two-magnon states
but also the continua of multi-magnon states.
Some properties of multi-magnon continua 
were examined in \cite{bib1}.
More recently,
we have noted that the spin-$\frac{1}{2}$ transverse $XX$ chain,
which can be mapped via the Jordan-Wigner transformation onto noninteracting spinless fermions,
is a model
whose dynamics is governed by continua of multi-fermion excitations.
In particular,
the dynamic trimer structure factor 
involves two-fermion and four-fermion excitations \cite{bib2}.
In the present report we compare and contrast 
the general and specific properties 
of the four-fermion excitation continuum,
which contributes to the dynamics of trimer fluctuations.

To be specific,
we consider the spin-$\frac{1}{2}$ transverse $XX$ chain 
with the Hamiltonian
\begin{eqnarray}
H=\sum_{n}J\left(s_n^xs_{n+1}^x+s_n^ys_{n+1}^y\right)
+\sum_{n}\Omega s_n^z.
\label{1}
\end{eqnarray}
We will set further $J=-1$.
The trimer operator is defined as
$T_n=s_{n}^xs_{n+2}^x+s_{n}^ys_{n+2}^y$
and
the corresponding dynamic structure factor
\begin{eqnarray}
S_{TT}(\kappa,\omega)
=\sum_l{\rm{e}}^{-{\rm{i}}\kappa l}
\int_{-\infty}^{\infty}{\rm{d}}t{\rm{e}}^{{\rm{i}}\omega t}
\langle \Delta T_{n}(t) \Delta T_{n+l}(0) \rangle,
\label{2}
\end{eqnarray}
$\Delta T_n(t)=T_{n}(t)-\langle T\rangle$
can be written as a sum of the two-fermion contribution
$S_{TT}^{(2)}(\kappa,\omega)$
and the four-fermion contribution
$S_{TT}^{(4)}(\kappa,\omega)$
with
\begin{eqnarray}
S_{TT}^{(2)}(\kappa,\omega)
=\int{\rm{d}}\kappa_1{\rm{d}}\kappa_2
C^{(2)}(\kappa_1,\kappa_2)
n_{\kappa_1}(1-n_{\kappa_2})
\nonumber\\
\cdot
\delta(\omega+\Lambda_{\kappa_1}-\Lambda_{\kappa_2})
\delta_{\kappa+\kappa_1-\kappa_2,0},
\label{3}
\\
S_{TT}^{(4)}(\kappa,\omega)
=\frac{1}{4\pi^2}
\int{\rm{d}}\kappa_1\ldots{\rm{d}}\kappa_4
C^{(4)}(\kappa_1,\ldots,\kappa_4)
\nonumber\\
\cdot
n_{\kappa_1}n_{\kappa_2}
(1-n_{\kappa_3})(1-n_{\kappa_4})
\nonumber\\
\cdot
\delta(\omega+\Lambda_{\kappa_1}+\Lambda_{\kappa_2}-\Lambda_{\kappa_3}-\Lambda_{\kappa_4})
\delta_{\kappa+\kappa_1+\kappa_2-\kappa_3-\kappa_4,0}.
\label{4}
\end{eqnarray}
Here
$C^{(2)}(\kappa_1,\kappa_2)$,
$C^{(4)}(\kappa_1,\ldots,\kappa_4)\ge 0$
are certain functions the explicit expressions for which are given in \cite{bib2},
$n_\kappa=(1+\exp(\beta\Lambda_\kappa))^{-1}$
is the Fermi function,
$\Lambda_\kappa=\Omega+J\cos\kappa$,
and $-\pi\le\kappa<\pi$ is the quasi-momentum 
which parameterizes the Jordan-Wigner fermions.
It is easy to note
that Eq. (\ref{3}) coincides with 
the dynamic $zz$ structure factor 
if $C^{(2)}(\kappa_1,\kappa_2)=1$
or with the dynamic dimer structure factor 
if $C^{(2)}(\kappa_1,\kappa_2)=\cos^2\frac{\kappa_1+\kappa_2}{2}$
(see \cite{bib2}).
These dynamic quantities are governed exclusively 
by the two-fermion 
(one particle and one hole)
excitations.
The properties of the two-fermion excitation continuum were examined in \cite{bib3,bib4}.

In contrast,
Eq. (\ref{4}) is governed exclusively by the four-fermion 
(two particles and two holes)
excitation continuum the properties of which are described concisely below.
The specific features of the four-fermion contribution to 
$S_{TT}(\kappa,\omega)$ (\ref{2})
are controlled by the function $C^{(4)}(\kappa_1,\ldots,\kappa_4)$.
To display the generic behavior of a 
four-fermion dynamic quantity 
we also consider Eq. (\ref{4})  
with $C^{(4)}(\kappa_1,\ldots,\kappa_4)=1$
(compare Figs. 1 and 2).
\begin{figure}[!ht]
\begin{center}
\includegraphics[width=0.45\textwidth]{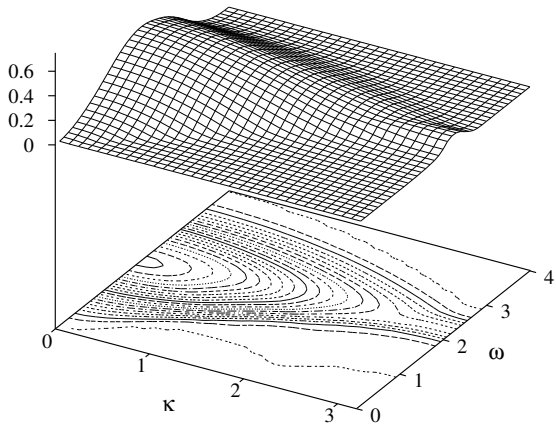} 
\end{center}
\vspace{0mm}
\caption{$S_{TT}^{(4)}(\kappa,\omega)$ (\ref{4}) 
for the chain (\ref{1}) with $J=-1$,
$\Omega=0.25$ at zero temperature ($\beta\to\infty$).}
\label{fig1}
\end{figure}
\begin{figure}[!ht]
\begin{center}
\includegraphics[width=0.45\textwidth]{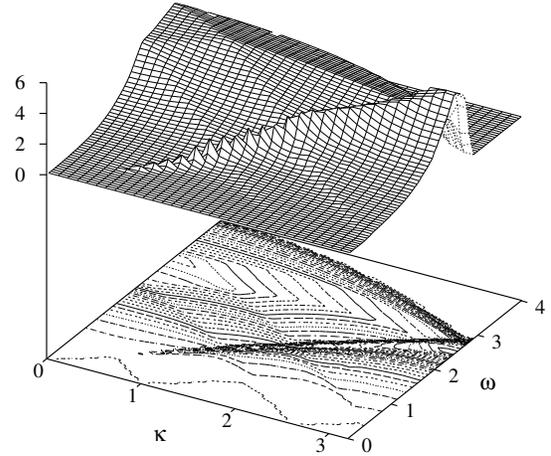} 
\end{center}
\vspace{0mm}
\caption{The same as in Fig. 1 but with $C^{(4)}(\kappa_1,\ldots,\kappa_4)=1$.}
\label{fig2}
\end{figure}

The four-fermion dynamic quantity can have nonzero values 
only in a restricted region of the $\kappa$--$\omega$ plane. 
At nonzero temperatures one immediately finds the upper boundary 
of the four-fermion excitation continuum,
$4\vert J\vert\cos\frac{\kappa}{4}$.
At zero temperature 
the Fermi functions in (\ref{4}) come into play 
and both the upper and the lower boundaries of the four-fermion excitation continuum 
become complicated $\Omega$-dependent functions of $\kappa$.
For $\Omega=0.25$ the upper boundary remains $4\vert J\vert\cos\frac{\kappa}{4}$,
whereas the lower boundary
 assumes 
the following values as $\kappa$ increases from 0 to $\pi$:
$\omega_l^{(1)}=2\vert J\vert\sin\frac{\kappa}{2}\sin(\alpha-\frac{\kappa}{2})$,
$\omega_l^{(2)}=4\vert J\vert\cos\frac{\kappa}{4}\cos(\alpha+\frac{\kappa}{4})$,
$\omega_l^{(3)}=-2\vert J\vert\sin(\alpha+\frac{\kappa}{2})\sin(2\alpha+\frac{\kappa}{2})$,
$\omega_l^{(1)}$,
$\omega_l^{(4)}=-2\vert J\vert\sin(\alpha-\frac{\kappa}{2})\sin(2\alpha-\frac{\kappa}{2})$
with $\cos\alpha=\frac{\Omega}{\vert J\vert}$.
To find these boundaries 
we (numerically) seek for
the extrema of $\cos\kappa_1+\cos\kappa_2-\cos\kappa_3-\cos\kappa_4$ 
with the restrictions imposed by the Fermi functions 
(see (\ref{4}))
for $0\le\kappa\le\pi$
and determine 
the values of $\kappa_1,\ldots,\kappa_4$
at which such extrema occur.
We find simple relations between the quantities 
$\kappa_1,\ldots,\kappa_4$ and $\kappa$, $\alpha$ 
obtaining as a result the upper and the lower boundaries 
of the four-fermion excitation continuum.

The four-fermion dynamic quantities may exhibit Van Hove cusp singularities 
akin to the three dimensional density of states.
These singularities occur along the lines 
$2\vert J\vert\sin\frac{\kappa}{2}$,
$4\vert J\vert\sin\frac{\kappa}{4}$,
and
$4\vert J\vert\cos\frac{\kappa}{4}$.

Comparing Figs. 1 and 2 we see 
how some characteristic features of the four-fermion excitation continuum 
are smeared out owing to $C^{(4)}(\kappa_1,\ldots,\kappa_4)\ne 1$.

Finally, 
we note that spin-$\frac{1}{2}$ $XX$ chains
are realized in some quasi-one-dimensional magnetic insulators
(e.g. such as Cs$_2$CoCl$_4$ \cite{bib5}).
The dynamic dimer structure factor 
is relevant to phonon-assisted optical adsorption \cite{bib6,bib7};
the direct experimental relevance of the dynamic trimer structure factor is less evident.
However,
our results may be important from the theoretical point of view
since the four-fermion dynamic trimer structure factor 
is a quantity of intermediate complexity 
between the two-fermion dynamic $zz$ structure factor 
and the multi-fermion dynamic $xx$ ($yy$) structure factor.

This study was supported by the STCU under the project \#1673.


\end{document}